# The Radio Structure of the γ-Ray Narrow-line Seyfert 1 Galaxy SDSS J211852.96-073227.5

Xi Shao[1,2,3] , Minfeng Gu[1,3] , Yongjun Chen[3,4] , Hui Yang[5] , Su Yao[6] , Weimin Yuan[7], and Zhiqiang Shen[3,4]
[1] Key Laboratory for Research in Galaxies and Cosmology, Shanghai Astronomical Observatory, Chinese Academy of Sciences, 80 Nandan Road, Shanghai 200030, People's Republic of China; gumf@shao.ac.cn, shaoxi@shao.ac.cn
[2] University of Chinese Academy of Sciences, 19A Yuquan Road, Beijing 100049, People's Republic of China
[3] Shanghai Astronomical Observatory, Chinese Academy of Sciences, Shanghai 200030, People's Republic of China
[4] Key Laboratory of Radio Astronomy, Chinese Academy of Sciences, 2 West Beijing Road, Nanjing, Jiangsu 210008, People's Republic of China
[5] Department of Physics, The George Washington University, 725 21st Street NW, Washington, DC 20052, USA
[6] Max-Planck-Institut für Radioastronomie, Auf dem Hügel 69, D-53121 Bonn, Germany
[7] Key Lab for Space Astronomy and Technology, National Astronomical Observatories, Chinese Academy of Sciences, Beijing 100012, People's Republic of China
*Received 2022 August 2; revised 2022 November 13; accepted 2022 November 28; published 2023 February 2*

## Abstract

The γ-ray narrow-line Seyfert 1 (NLS1) galaxies can be considered to be the third class of γ-ray active galactic nuclei possessing relativistic jets. In this paper, we present multi-band high-resolution Very Long Baseline Array (VLBA) images of the γ-ray NLS1, SDSS J211852.96-073227.5 (J2118-0732, $z = 0.26$). We find a core-jet radio morphology and significant flux density variations in the radio core. The high brightness temperature estimated from VLBA images and core variability demonstrate that it exhibits substantial relativistic beaming effects. By considering radio emission in several bands, we find that the source has an inverted spectrum above 1 GHz but a steep spectrum at low frequencies ranging from 74 MHz–1 GHz; these may arise from the present activity and the old diffuse/extended emission, respectively. The core-jet morphology, significant flux density variations, and beaming effect make J2118-0732 resemble a blazar. Considering the low mass of its central black hole and the ongoing merger environment, J2118-0732 may represent a low-mass, low-power counterpart of blazars, and may finally evolve into a blazar.

*Unified Astronomy Thesaurus concepts:* Radio jets (1347); Active galactic nuclei (16); Seyfert galaxies (1447); Radio cores (1341)

## 1. Introduction

Narrow-line Seyfert 1 (NLS1) galaxies, as one subset of active galactic nuclei (AGNs), have permitted emission lines that are narrow (defined as FWHM(H$\beta$) < 2000 km s$^{-1}$; Goodrich 1989) and are characterized by relatively strong Fe II multiplets and weak [O III] emission [O III]/H$\beta$ flux ratio <3) (Osterbrock & Pogge 1985). NLS1s are more preferentially radio quiet compared to the whole AGN population, and only 7% of them are radio-loud (Zhou et al. 2006; Rakshit et al. 2017), while this fraction is 15% in quasars and normal Seyfert 1 galaxies (Ivezić et al. 2002). Together with radio-quiet NLS1s, RLNLS1s have lower black hole masses and higher Eddington ratios compared to general type 1 AGN populations (e.g., Yuan et al. 2008), thus reflecting that they are likely at an early stage of evolution (see review by Komossa 2008). A majority of RLNLS1s were found to be hosted in spiral galaxies (Zhou et al. 2007; Olguín-Iglesias et al. 2017) with relatively intense star formation activity (Caccianiga et al. 2015). These features are significantly different from typical radio-loud AGNs, including blazars, which are usually hosted in elliptical galaxies (e.g., Olguín-Iglesias et al. 2016).

Evidence that indicates the presence of relativistic jets in some RLNLS1s has been gradually accumulated (Komossa et al. 2006), especially at the highest radio-loudness regime, due to their blazar-like characteristics (Zhou et al. 2007; Yuan et al. 2008). Very Long Baseline Interferometry (VLBI) observations provide a direct way to detect the relativistic jet and explore its properties. The study of a sample of RLNLS1s shows that they usually have a compact radio morphology with a one-sided core-jet structure detected at the parsec scale (e.g., Gu et al. 2015). With the successful detection of γ-ray emission from some NLS1s by the Large Area Telescope (LAT) on board the Fermi Satellite (Abdo et al. 2010), these γ-ray emitting RLNLS1s became a new class of galaxies with relativistic jets emitting at γ-ray energies in addition to blazars and some other radio galaxies (Hartman et al. 1999).

There are around 20 confirmed γ-ray NLS1s (Romano et al. 2018; Foschini et al. 2021). Nine of them are included in the Fourth Fermi-LAT source catalog (4FGL; Abdollahi et al. 2020), most of which present a one-sided jet emerging from the radio core. Five sources have been monitored with the Monitoring Of Jets in Active galactic nuclei with VLBA Experiments (MOJAVE) program (Lister et al. 2016), with apparent superluminal motion detected in three NLS1s (1H 0323342, SBS 0846513, PMN J09480022), and subluminal motion in two others (SDSS J12220413, PKS 1502036). Besides these nine bona fide NLS1s, 10 additional sources have radio detection, with most of them being detected the radio emission on the kiloparsec scale (Spencer et al. 1989; Helfand et al. 2015; Gu et al. 2015; Lister et al. 2016; Berton et al. 2018, 2020; Singh & Chand 2018), nearly half of them have one-sided parsec-scale core-jet structures (Linford et al. 2012; Orienti et al. 2012; D'Ammando et al. 2012; Lister et al. 2016; An et al. 2017). It has been proposed that γ-ray NLS1s could be regarded as low-luminosity counterparts of powerful blazars (especially flat-spectrum radio quasars (FSRQs); Berton et al. 2016) since they are at the low end of the disk-jet connection for blazars (Foschini et al. 2015; Paliya et al. 2019) and γ-ray NLS1s are likely to be young, low-power, precursors of FSRQs (Paliya 2019).

The target SDSS J211852.96-073227.5 (J2118-0732 hereafter) is a γ-ray NLS1 at a redshift of 0.26, as found by Yang et al. (2018) and Paliya et al. (2018). Interestingly, J2118-0732 is hosted in a merging system and its companion galaxies both have pseudo-bulges and are late-type galaxies (Järvelä et al. 2020). Yang et al. (2018) studied this source with multifrequency observations and found rapid variations (intraday) behavior in the infrared band, a flat X-ray spectrum with a flux variability factor around 2.5 over the course of 5 months, and incidents of γ-ray flares during 2009–2013. All these properties are similar to those in blazars. Without being affected by the dust obscuration, high-resolution VLBI radio observations provide direct images of jets at the parsec scale, and are an excellent tool to study the radio morphology and jet properties. VLBI further enables the study of the nature of the source, especially allowing for the investigation of the relationship between jet formation and the merger process (e.g., Yang et al. 2018; Järvelä et al. 2020; Paliya et al. 2020).

In this paper, we present a study of the jet properties of J2118-0732 by using multifrequency VLBA radio observations. The observations and data reduction are discussed in Section 2, while Section 3 presents our results. A discussion and our conclusions are given in Sections 4 and 5, respectively. Throughout the paper, the cosmological parameters of $H_0 = 70 \text{ km s}^{-1} \text{ Mpc}^{-1}$, $\Omega_m = 0.3$, and $\Omega_\lambda = 0.7$ are used. The radio spectral index $\alpha$ is defined as $S_\nu \propto \nu^\alpha$ with the radio flux density $S_\nu$ at frequency $\nu$.

## 2. Observation and Data Reduction

For our analysis, we utilized our new VLBA observations (Project code: BG252), archival public VLBA, and Very Large Array (VLA) data to study the morphology and radio properties of J2118−0732. Our VLBA observations were carried out in the C, S/X, and Ku bands (5.0, 2.3/8.4, and 15.1 GHz) on 2018 March 30. The FWHM synthesized beam sizes of the S, C, X, and Ku bands are $10.20 \times 3.68$ mas ($41.0 \times 14.8$ pc), $3.91 \times 1.64$ mas ($15.7 \times 6.6$ pc), $2.33 \times 0.99$ mas ($9.4 \times 4.0$ pc), and $1.37 \times 0.56$ mas ($5.5 \times 2.2$ pc), respectively. The total observing time was 7 hr, including a 60 minute integrated on-source time in each band, the antenna slewing time, and the time on the calibrator, BL Lac.

VLBA data reduction was carried out in a standard way with the National Radio Astronomy Observatory (NRAO) AIPS package (Greisen 2003). BL Lac was used as the calibrator to perform the amplitude self-calibration and bandpass calibration. Due to the time and frequency-dependent effects, bad data was flagged. Then self-calibration was performed for the source. The final uv visibility was produced under natural weight with DIFMAP (Shepherd 1997). The task CLEAN was performed to map the true structure in the concatenated image. In the task MODELFIT, we modeled the brightness distribution of bright knots in the uv-visibility data with circular Gaussian components, from which the measurements of radio components were obtained. The measurement results from the Gaussian fits, such as flux density, position, size, and axis ratio, are shown in Table 1, and the parsec-scale radio images are exhibited in Figure 1.

We collected the VLBA archival data from the Astrogeo Center[8] for the observations of J2118-0732 made at 4.3 and 7.6 GHz on 2016 June 5 and 2018 May 30. We directly downloaded the FITS files to produce the radio images with DIFMAP. For our VLBA observations, we used Gaussians to fit the radio components. The images from the archival data are displayed in Figure 2, and the measurements of radio components are shown in Table 2. To study the kiloparsec-scale radio structure, we used the archival VLA[9] data for J2118-0732 (Observation ID: AH640). These observations were at 8.4 GHz with an A configuration and were made on 1998 May 18. The FWHM synthesized beam size of the X band is $350 \times 224$ mas ($1.4 \times 0.9$ kpc). The reduced image is shown in Figure 3.

## 3. Results

### 3.1. The Radio Structure

Our VLBA images in Figure 1, show that J2118-0732 has a compact radio structure in all four bands, being slightly resolved in the C and Ku bands (panels (b) and (d)), while unresolved in the S and X bands (panels (a) and (c)), by visual inspection. The model-fit results performed with DIFMAP show two radio components in the S, X, and Ku bands, while showing three components in the C band (see Table 1). The function STDEV was the error estimate given by the process of the task MODELFIT. The brightest component in each band is identified as the radio core as it has a flat, or even inverted spectrum, from 2.3–15 GHz, i.e., the spectral index $\alpha \geqslant -0.5$. Thus, the source exhibits a core-jet structure, albeit with very weak jet emission.

The images from two epoch VLBA observations on 2016 June 5 and 2018 May 30 are exhibited in Figure 2. Upon inspection, it appears that J2118-0732 only shows a compact core in the C and X bands in 2016, while the target has a core-jet structure in both bands on 2018 May 30. The radio core is again identified as the brightest component with inverted spectrum (see Table 2).

The VLA image of J2118-0732 observed at 8.46 GHz on 1998 May 18 (see Figure 3) only shows an unresolved core with a flux density of 91.3 mJy. The core flux density is higher than the total VLBA flux density in the X band, by at least 15 mJy (for the highest VLBA flux density measured on 2018 May 30). This indicates that at least 16% of the VLA flux from the kiloparsec scale could be resolved out at parsec-scale resolution. Alternatively, this may be due to the flux density variations (see the next section).

Based on the VLBI images, the brightness temperatures, $T_B$, of the radio core can be estimated in different bands in the rest frame following the formula of Shen et al. (1997):

$$T_B = 1.22 \times 10^{12}(1+z)\frac{S_m}{\nu_m^2 \theta_d^2}, \quad (1)$$

where $z$ is the redshift, $S_m$ and $\nu_m$ are the flux density (in units of jansky) and observing frequency (in units of gigahertz), and $\theta_d = \sqrt{ab}$ presents the angular diameter of the radio core with $a$ and $b$ being the major and minor axes in units of milliarcseconds, respectively. As shown in Tables 1 and 2, the radio cores in four bands all show high brightness temperatures, with $T_B$ ranging from $10^{10.9}$ to $10^{11.8}$ K. Such high brightness temperatures strongly indicate that the radio emission originates from a relativistic jet. With the brightness

---

[8] http://astrogeo.org/

[9] http://data.nrao.edu/

Table 1
Results Obtained from the VLBA Images of J2118-0732 Observed on 2018 March 30

| Bands | Comps. | $S$ (mJy) | $\sigma_S$ (mJy) | $r$ (mas) | $\sigma_r$ (mas) | $\theta$ (degree) | $a$ (mas) | $\sigma_a$ (mas) | $b/a$ | $\log T_B$ (K) | $\delta_1$ | $\delta_2$ | $\alpha$ |
|---|---|---|---|---|---|---|---|---|---|---|---|---|---|
| (1) | (2) | (3) | (4) | (5) | (6) | (7) | (8) | (9) | (10) | (11) | (12) | (13) | (14) |
| $S$ | Core | 42.67 | 0.47 | 0.121 | 0.024 | −0.1 | 0.38 | 0.08 | 1.00 | 10.9 | 1.7 | 0.1 | ⋯ |
|   |   | 3.24 | 0.48 | 4.136 | 0.374 | −169.9 | 1.45 | 0.33 | 1.00 | ⋯ | ⋯ | ⋯ | −0.48 |
| $C$ | Core | 29.41 | 0.05 | 0.066 | 0.002 | 1.7 | 0.10 | 0.01 | 1.00 | 11.3 | 3.8 | 0.2 | ⋯ |
|   |   | 1.39 | 0.05 | 2.373 | 0.052 | −173.2 | 0.07 | 0.46 | 1.00 | ⋯ | ⋯ | ⋯ | 0.53 |
|   |   | 0.94 | 0.05 | 4.833 | 0.073 | −168.6 | 1.23 | 0.08 | 1.00 | ⋯ | ⋯ | ⋯ | ⋯ |
| $X$ | Core | 38.60 | 0.11 | 0.006 | 0.002 | 172.4 | 0.07 | 0.02 | 1.00 | 11.2 | 3.1 | 0.2 | ⋯ |
|   |   | 1.89 | 0.11 | 1.925 | 0.046 | −174.7 | 0.30 | 0.10 | 1.00 | ⋯ | ⋯ | ⋯ | 0.42 |
| $Ku$ | Core | 49.44 | 0.08 | 0.002 | 0.001 | −1.2 | 0.02 | 0.01 | 1.00 | 11.7 | 10.8 | 0.5 | ⋯ |
|   |   | 1.13 | 0.09 | 1.266 | 0.061 | −171.9 | 0.58 | 0.07 | 1.00 | ⋯ | ⋯ | ⋯ | ⋯ |

**Note.** Column (1): observing bands; Column (2): radio components; Columns (3) and (4): flux density of the radio components and its standard deviation; Columns (5)–(7): component position, its standard deviation, and position angle; Columns (8) and (9): major axis and its standard deviation; Column (10): axial ratio; Column (11): brightness temperature; Columns (12) and (13): Doppler factor calculated for the intrinsic brightness temperatures of $5 \times 10^{10}$ and $10^{12}$ K, respectively; Column (14): spectral index derived from the adjacent bands, $\alpha = \log(S_1 - S_2)/\log(\nu_1 - \nu_2)$, where $S_1$ and $S_2$ are the flux densities of cores at frequencies of $\nu_1$ and $\nu_2$, respectively. For example, −0.48 is the spectral index between the $S$ and $C$ bands.

temperature, the Doppler factor $\delta$, an indicator of the beaming effect of a relativistic jet, can be constrained by

$$\delta = T_B / T'_B, \quad (2)$$

where $T'_B$ is the intrinsic brightness temperature, usually taken as $5 \times 10^{10}$ K under the equipartition condition (Readhead 1994) or but possibly as high as $10^{12}$ K in the case of the so-called inverse Compton catastrophe (Kellermann & Pauliny-Toth 1969). These alternative results for $\delta$ are shown in Tables 1 and 2, while the first set is shown in Figure 4. Significant beaming effects are found, with the Doppler factor ranging from 1.7–12.4 by taking the equipartition brightness temperature as the limiting intrinsic value. The Doppler factor varies at different frequencies, and even at different epochs in a single band. However, the brightness temperature is no larger than $10^{12}$ K, indicating that the beaming effect in the source is not extreme compared to that in blazars, in which extreme brightness temperatures above $10^{12}$ K are frequently observed (e.g., Kovalev et al. 2009)

### 3.2. Variability

From Tables 1 and 2, the presence of flux density variations in the radio core can be clearly seen; these are consistent with changes in the effect of beaming as indicated by the measured brightness temperature, which could possibly be caused by either the geometry effect (e.g., the change of jet direction) or a physical reason (e.g., jet acceleration from new jet ejection). We take the fractional amplitude $|\Delta S|/\langle S \rangle$ to quantify the radio variability, where $\Delta S = S_1 - S_2$ is the flux difference at two epochs, $\langle S \rangle$ presents the average flux density at two epochs, and the significance of flux density variability can be estimated by $\sigma = |\Delta S|/(\sigma_{S_1}^2 + \sigma_{S_2}^2)^{1/2}$ (Yuan et al. 2008). The results are exhibited in Table 3. When comparing 2016 June with 2018 March, the flux density in the $C$ band slightly increases ($|\Delta S|/\langle S \rangle = 9.5\%$), whereas that in the $X$ band decreases with $|\Delta S|/\langle S \rangle = 16\%$. In contrast, much larger variations were found during the short interval between 2018 March and 2018 May when the flux density of the source significantly rose in the same proportion in both the $C$ band ($|\Delta S|/\langle S \rangle = 64.4\%$) and the $X$ band ($|\Delta S|/\langle S \rangle = 64.8\%$).

To further investigate the radio variability, we calculated the variability in brightness temperature $T_B^*$ (Ghosh & Punsly 2007). This differs from $T_B$, which is calculated from the measurements based on the model fitting on radio images, as $T_B^*$ is based on the flux density variability and the light-crossing time through the variable part,

$$T_B^* = \frac{2D_L^2 |\Delta S_\nu|}{\pi k (1+z) \nu^2 (\Delta t)^2} = 5.88 \times 10^{10} (1+z)^{-1}$$
$$\times \left(\frac{D_L}{\text{Mpc}}\right)^2 \left(\frac{|\Delta S_\nu|}{\text{mJy}}\right) \left(\frac{\nu}{\text{GHz}}\right)^{-2} \left(\frac{\Delta t}{\text{day}}\right)^{-2} \text{K}, \quad (3)$$

where $k$ is the Boltzmann constant, $z$ is the source redshift, $\Delta S_\nu$ represents the flux density change in the period $\Delta t$ at frequency $\nu$, and the luminosity distance $D_L = \frac{c(1+z)}{H_0} \int_0^z [\Omega_\lambda + \Omega_m(1+x^3)]^{-1/2} dx$, where $c$ is the speed of light. The Doppler factor can be estimated from the variability in brightness temperature as $\delta^* = (T_B^*/T_B')^{1/3}$ by taking $T_B'$ as either $5 \times 10^{10}$ or $10^{12}$ K. The results are given in Table 3. Clearly, the highest variability in brightness temperature beyond $10^{13}$ K occurs between 2018 March and 2018 May, and the results from the largest flux density variations are seen in the shortest measured time duration. Values for $T_B^*$ exceeding even the inverse Compton limit strongly imply an origin from the beaming effect in the jet. The calculated Doppler factors are modest and generally consistent with the results obtained from the VLBA images.

### 3.3. Jet Power

As the emission of the radio core can be used as a good proxy of jet power (Blandford & Königl 1979), we estimated the jet power following the empirical formulas in Foschini (2014), which were based on a sample of AGNs, including NLS1s,

$$\log P_{\text{jet,rad}} = (12 \pm 2) + (0.75 \pm 0.04) \log L_{15\text{GHz}}, \quad (4)$$

$$\log P_{\text{jet,kin}} = (6 \pm 2) + (0.90 \pm 0.04) \log L_{15\text{GHz}}, \quad (5)$$

where $P_{\text{jet,rad}}$ and $P_{\text{jet,kin}}$ are the jet radiative and kinetic power, respectively. The luminosity of the radio core at 15 GHz is

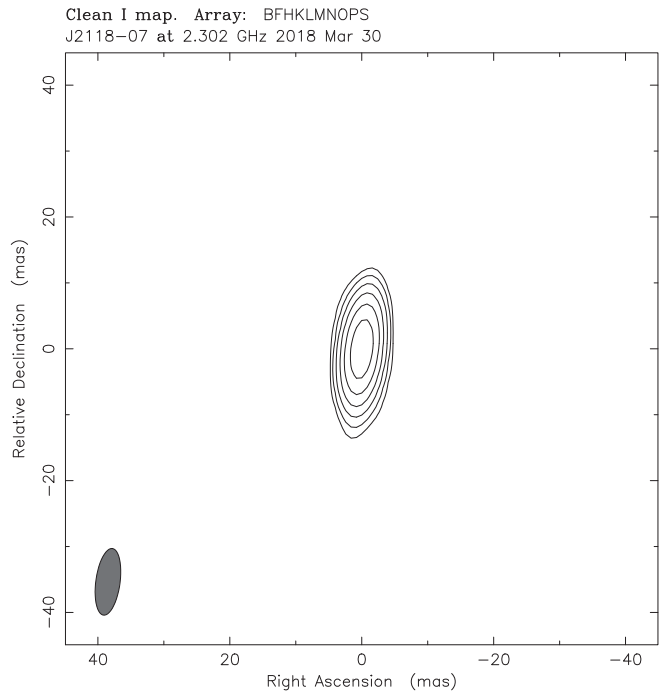
(a) S band

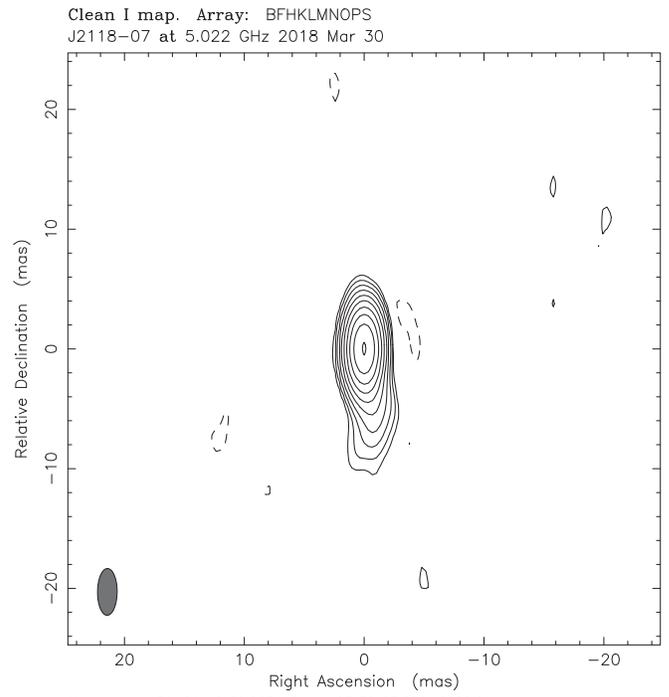
(b) C band

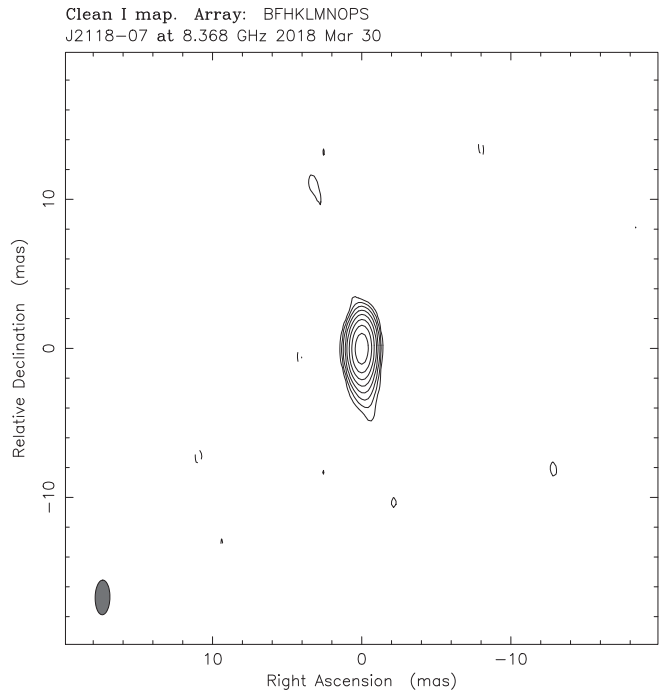
(c) X band

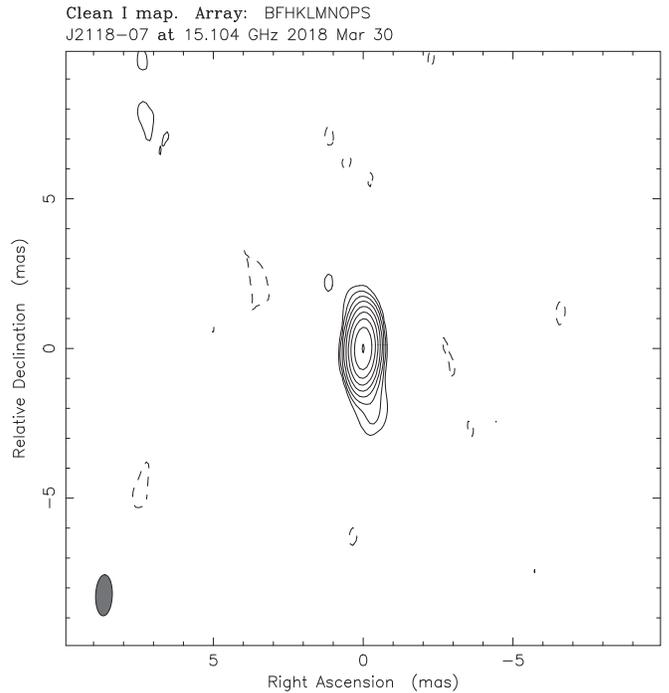
(d) Ku band

**Figure 1.** Our VLBA images of J2118-0732 observed on 2018 March 30. (a) *S* band, (b) *C* band, (c) *X* band, and (d) *Ku* band. The beam FWHMs from panels (a)–(d) are 41.0 × 14.8, 15.7 × 6.6, 9.4 × 4.0, and 5.5 × 2.2 pc, respectively.

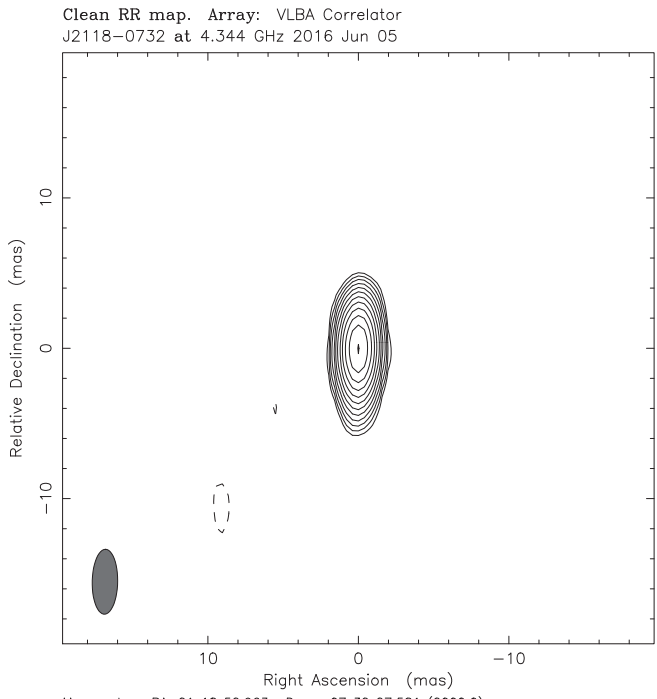
(a) C band

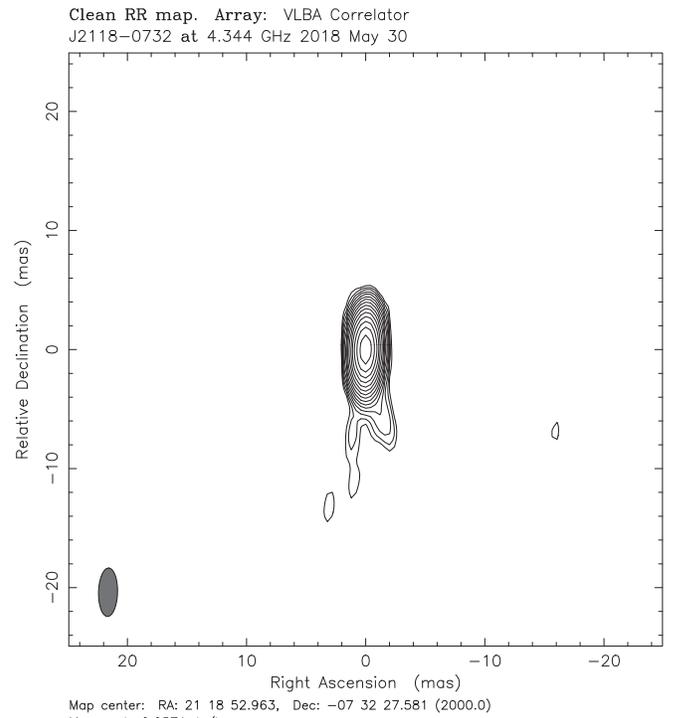
(b) C band

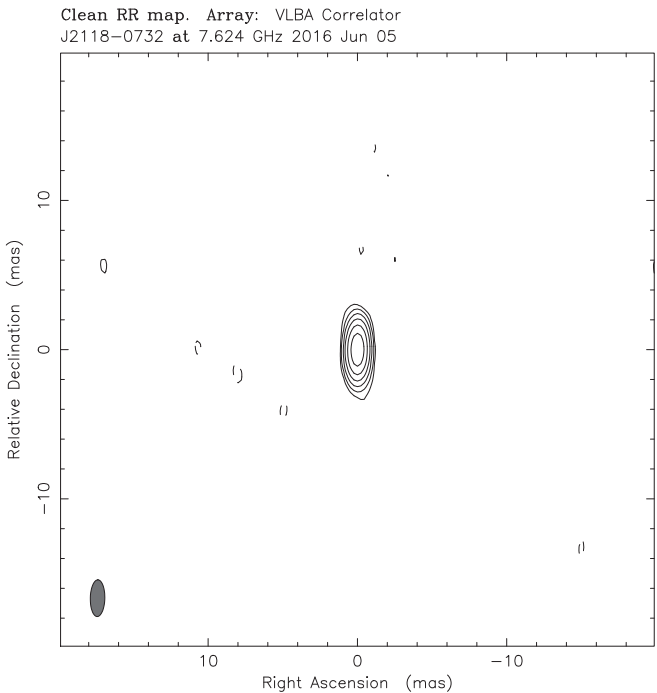
(c) X band

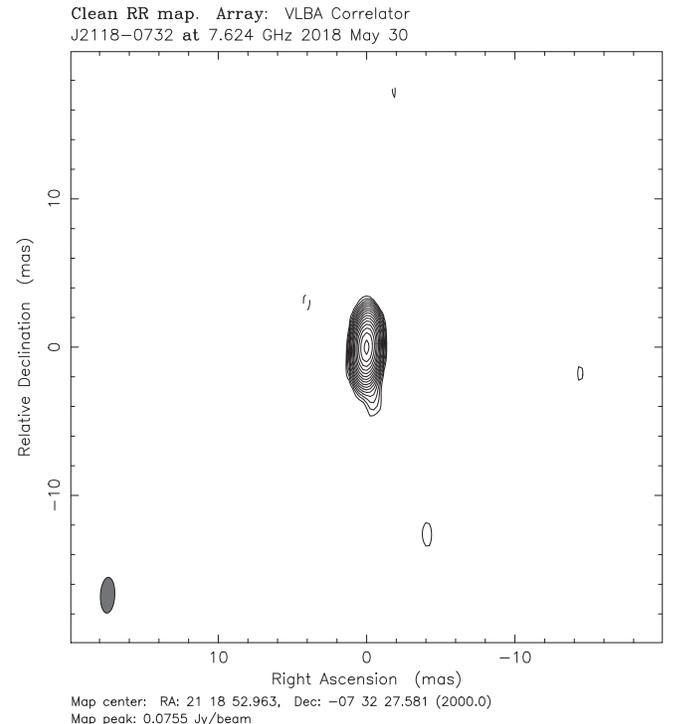
(d) X band

**Figure 2.** VLBA images of J2118-0732 at the *C* and *X* bands, from the data obtained from the Astrogeo Center: (a) and (c) on 2016 June 5; (b) and (d) on 2018 May 30. The beam FWHMs from panels (a)–(d) are 17.4 × 6.8, 16.4 × 6.4, 9.9 × 3.9, and 9.7 × 3.9 pc, respectively.

Table 2
Results Obtained from the VLBA Images of J2118-0732 Observed on 2016 June 5 and 2018 May 30

| Bands | Comps. | $S$ (mJy) | $\sigma_S$ (mJy) | $r$ (mas) | $\sigma_r$ (mas) | $\theta$ (degree) | $a$ (mas) | $\sigma_a$ (mas) | $b/a$ | $\log T_B$ (K) | $\delta_1$ | $\delta_2$ |
|---|---|---|---|---|---|---|---|---|---|---|---|---|
| (1) | (2) | (3) | (4) | (5) | (6) | (7) | (8) | (9) | (10) | (11) | (12) | (13) |
| $C^a$ | Core | 26.74 | 0.58 | 0.008 | 0.024 | −135.4 | 0.06 | 0.44 | 1.00 | 11.8 | 12.4 | 0.6 |
|  |  | 1.30 | 0.61 | 3.566 | 0.820 | −179.0 | 0.72 | 1.10 | 1.00 | ⋯ | ⋯ | ⋯ |
| $C^b$ | Core | 57.36 | 0.24 | 0.004 | 0.005 | −9.4 | 0.13 | 0.10 | 1.00 | 11.4 | 5.2 | 0.3 |
|  |  | 1.10 | 0.22 | 5.968 | 0.263 | −165.4 | 0.19 | 1.20 | 1.00 | ⋯ | ⋯ | ⋯ |
|  |  | 0.78 | 0.24 | 7.338 | 0.454 | 170.4 | 0.36 | 25.0 | 1.00 | ⋯ | ⋯ | ⋯ |
|  |  | 0.55 | 0.23 | 13.944 | 0.568 | 167.3 | 0.20 | 1.20 | 1.00 | ⋯ | ⋯ | ⋯ |
| $X^a$ | Core | 45.29 | 0.44 | 0.006 | 0.008 | −169.7 | 0.11 | 0.05 | 1.00 | 11.0 | 2.0 | 0.1 |
| $X^b$ | Core | 75.57 | 0.20 | 0.004 | 0.002 | −179.2 | 0.06 | 0.02 | 1.00 | 11.7 | 9.9 | 0.5 |
|  |  | 0.88 | 0.20 | 2.910 | 0.189 | −173.7 | 0.02 | 5.90 | 1.00 | ⋯ | ⋯ | ⋯ |

**Note.** Column (1): observing bands, superscript $^a$ — 2016 June 5 and superscript $^b$ — 2018 May 30; Column (2): L radio components; Columns (3) and (4): flux density of the radio components and its standard deviation; Columns (5)–(7): component position, its standard deviation, and position angle; Columns (8) and (9): major axis and its standard deviation; Column (10): axial ratio; Column (11): brightness temperature; Columns (12) and (13): Doppler factor calculated for the intrinsic brightness temperatures of $5 \times 10^{10}$ and $10^{12}$ K, respectively.

calculated by $L_{15\text{GHz}} = 4\pi D_L^2 \nu S_\nu (1+z)^{-(\alpha+1)}$, in which $S_\nu$ and $\alpha$ represent the radio core flux density and spectral index at 15 GHz, respectively. Here, we adopt $S_\nu = 49.67$ mJy (the Ku band on 2018 March 30) and $\alpha = 0.42$ between the X and Ku bands (see Table 1). Thus, we calculate $L_{15\text{GHz}} = 1.11 \times 10^{42}$ erg s$^{-1}$, and find that the jet radiative power and kinetic power are $P_{\text{jet,rad}} = 10^{43.5}$ and $P_{\text{jet,kin}} = 10^{43.8}$ erg s$^{-1}$, respectively. A discussion of the above results is given in Section 4.

## 4. Discussion

### 4.1. The Nature of J2118-0732

This target was previously classified as an NLS1 by various authors (Rakshit et al. 2017; Paliya et al. 2018; Yang et al. 2018), whereas it was reclassified as an intermediate-type galaxy by Järvelä et al. (2020), in which the broad component of the H$\beta$ line is modeled by a combination of two Gaussian profiles and the narrow component is modeled by a single Gaussian profile. The reclassification by Järvelä et al. (2020) was mainly based on the significant contribution of the narrow component to the H$\beta$ line and the decomposition of H$\beta$ into two components. However, the classification of NLS1 requires the FWHM of the overall broad H$\beta$ component, which was unfortunately not provided in their paper. On the other hand, the definition of an intermediate-type galaxy is not explicitly given in their work. For these reasons, we still take J2118-0732 as an NLS1 as in Rakshit et al. (2017), Yang et al. (2018), and Paliya et al. (2018). Komossa (2018) reported that J2118-0732 is in an interaction with a merging system. Of particular interest to us, this source was found to probably be a merging system according to Sloan Digital Sky Survey (SDSS) images in the r band, in which an extended structure was displayed in the northeastern direction (Yang et al. 2018). Further claimed as a rare nonlocal, interacting late-type Seyfert galaxy with relativistic jets (Järvelä et al. 2020; Paliya et al. 2020), J2118-0732 is a significant target to study the jet formation in late-type galaxies, and its relation to the merger system, in addition to two previous cases (1H 0323+342 (Antón et al. 2008), and FBQS J164442.5 +261913 (Olguín-Iglesias et al. 2017).

The unambiguous detection of the ongoing minor merger of J2118-0732 with a companion Seyfert 2 galaxy at a separation of ∼12 kpc has also been reported (Paliya et al. 2020). The authors proposed that the jet could be triggered by a galaxy merger as the merging timescale of 0.5−2 Gyr is considerably longer than its jet kinematic age (<15 kyr) estimated from the unresolved radio structure (Paliya et al. 2020). While our observations found the source to show a compact structure, some elongated and extended structure is clearly detected (see Figures 1 and 2). Indeed, the model fit on the VLBA images clearly shows jet components although they are rather weak compared to that of the radio core (see Tables 1 and 2). We found that the direction of the jet of J2118-0732 is likely perpendicular to the major axis of the pseudo-bulge and disk of its host galaxy. The jet position angles shown in the VLBA images are about −170° while the optical major axis is at about −75° to −80° (see Table 1 and Järvelä et al. 2020; Paliya et al. 2020). Previous studies of the samples of Seyfert galaxies suggest that the accretion disks are not aligned with the host galaxy disk (Ulvestad & Wilson 1984; Brindle et al. 1990; Baum et al. 1993; Schmitt et al. 1997; Clarke et al. 1998; Nagar & Wilson 1999; Schmitt et al. 2001), since no correlation between the position angles of the host galaxy major axis and that of the jets was found. Therefore, our source will be an interesting target to investigate the alignment between the accretion disk and host galaxy disk.

Having multi-epoch VLBA observations enables us to study the motion of the jet component, which is crucial to estimate the jet parameters, such as the bulk velocity and viewing angle. Here, we choose data in the X band for the study as the higher frequency provides higher resolution compared to the C band. There are useful observations at three epochs. Only a radio core was detected in the X band on 2016 June 5, while at least one jet component in addition to the radio core was found on 2018 March 30 and 2018 May 30 (see Tables 1 and 2). Based on the separation of the jet component from the core in 2018, we obtained an apparent velocity of $5.40c$ (5.91 mas yr$^{-1}$) for the detected jet component. The apparent velocity, in combination with the Doppler factor of 6.3 constrained from the variability (in Table 3), gives values of the jet bulk velocity of $0.98c$ at a

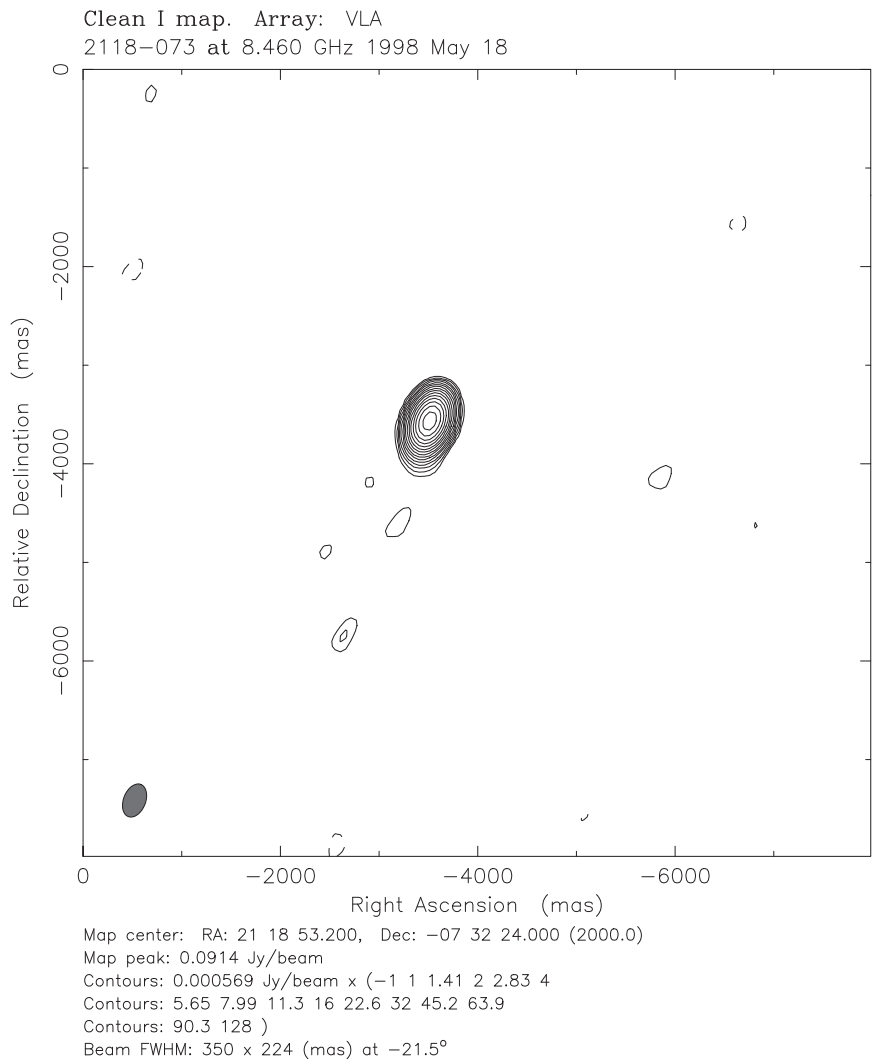

**Figure 3.** VLA image of J2118-0732 at the *X* band observed on 1998 May 18.

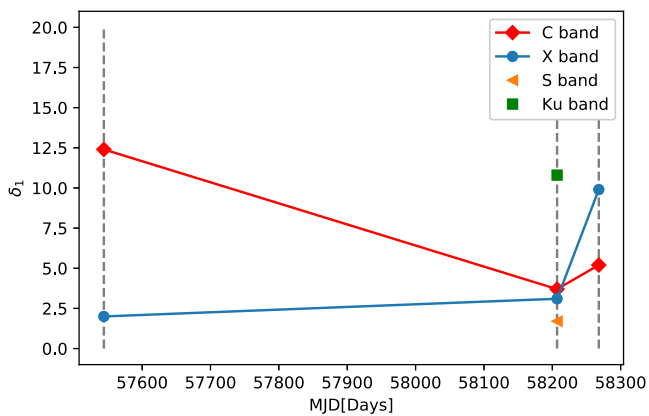

**Figure 4.** The variation in the Doppler factor (*diamonds*—*C* band, *circles*—*X* band, *triangles*—*S* band, *squares*—*Ku* band). The dashed lines from left to right represent the epochs of 2016 June 5, 2018 March 30, and 2018 May 30, respectively.

**Table 3**
Variability of J2118-0732 from VLBA observations

| Bands | $\Delta S$ (mJy) | $\frac{|\Delta S|}{\langle S \rangle}$ (%) | $\sigma$ | $\Delta t$ (day) | $\log T_B^*$ (K) | $\delta_1^*$ | $\delta_2^*$ |
|---|---|---|---|---|---|---|---|
| (1) | (2) | (3) | (4) | (5) | (6) | (7) | (8) |
| $C^b - C$ | 27.95 | 64.42 | 114.4 | 61 | 13.4 | 7.9 | 2.9 |
| $C - C^a$ | 2.67 | 9.51 | 4.6 | 663 | 10.3 | 0.7 | 0.3 |
| $X^b - X$ | 36.97 | 64.77 | 162.0 | 61 | 13.1 | 6.3 | 2.3 |
| $X - X^a$ | −6.69 | 15.95 | 14.8 | 663 | 10.3 | 0.7 | 0.3 |
| $C^b - C^a$ | 30.62 | 72.81 | 48.8 | 724 | 11.3 | 1.6 | 0.6 |
| $X^b - X^a$ | 30.28 | 50.11 | 62.7 | 724 | 10.9 | 1.1 | 0.4 |

**Note.** Column (1): related bands, superscript [a] — 2016 June 5 and superscript [b] — 2018 May 30; Column (2): the flux density difference between two epochs; Column (3): the fraction amplitude, where $<\Delta S>$ is the average of two flux densities; Column (4): the significance of the flux density variability; Column (5): the time separation between two observations; Column (6): the variability in brightness temperature; Columns (7) and (8): the variability in the Doppler factor calculated by $\delta^* = (T_B^*/T_B')^{1/3}$ with the intrinsic brightness temperatures of $5 \times 10^{10}$ and $10^{12}$ K, respectively.

viewing angle of 9°.0. These estimations however are tentative, since they are based on two-epoch observational results only, which need to be confirmed with more observations.

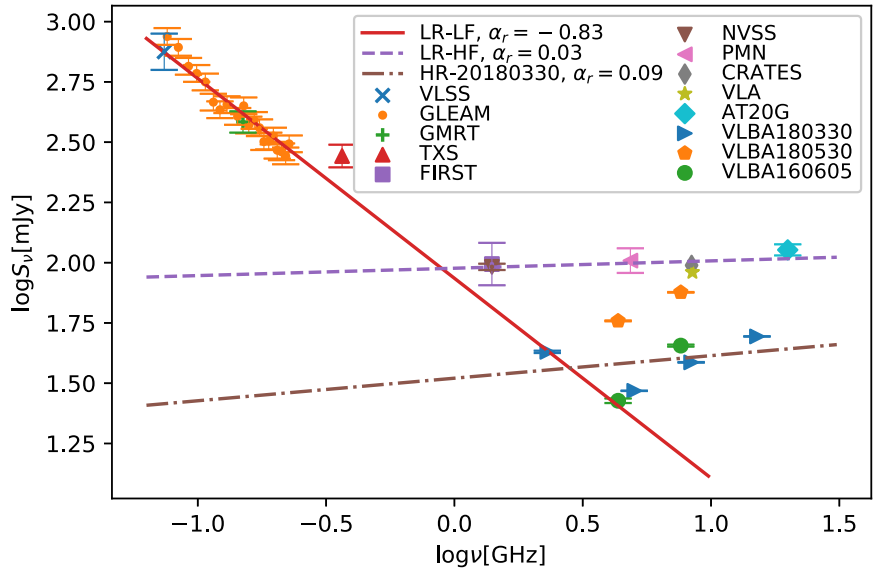

**Figure 5.** The radio spectrum of J2118-0732. The solid red line and purple dash line represent the power-law fit on the low-resolution data at low and high frequencies, respectively (see the text for details). The brown dotted–dashed line represents the power-law fit for the VLBA data obtained on 2018 March 30.

### 4.2. Radio Spectrum

Due to the flat spectrum below 4.8 GHz, J2118-0732 was regarded as a flat-spectrum source in the Combined Radio All-Sky Targeted Eight GHz Survey (CRATES; Healey et al. 2007). Later, J2118-0732 was found to exhibit a bending spectrum with $\alpha = -0.7$ between 74 MHz and 1.4 GHz, and a flatter spectrum above 1.4 GHz (Vollmer et al. 2010). By collecting low-resolution radio data, Yang et al. (2018) more recently obtained $\alpha_r = -0.66$ below 1.4 GHz, and $\alpha_r = 0.15$ above 1.4 GHz, using broken power-law fitting.

In Figure 5, we plot the radio spectrum of the source with the collected low-resolution multi-band data from VizieR Catalogs[10] (listed in Table 4), and our total VLBA flux densities. The frequency range of all these data spans 0.074–19.9 GHz and the low-resolution data clearly exhibit a concave spectrum. We fitted the low-resolution data with two power-law models, one below 1 GHz and the other above 1 GHz. We also find that the sub-gigahertz spectrum is steep, with $\alpha = -0.83$ while the higher-frequency range shows a barely inverted spectrum $\alpha = 0.03$. Figure 5 clearly shows that the spectra of the VLBA observations on 2016 June and 2018 May are all inverted between the C and X bands. With four-band measurements, however, the spectrum of our VLBA observations on 2018 March shows a concave shape with a transition around 5 GHz, with $\alpha = -0.48$ below and $\alpha = 0.53$ above (see Table 1). The relatively steep spectrum (but still flat) below 5 GHz may be partly caused by the lower resolution in the S band, which could include more extended emission and might be resolved out at higher frequencies. The power-law fit on an overall frequency range of 2.3–22 GHz gives an inverted spectrum with $\alpha = 0.09$ (see Table 5). To illustrate the general inverted spectrum, we fit the VLBA spectra with a power law for all data (i.e., data from 2016 in 2018), and the data in 2018, which is plotted in Figure 5 (see also Table 5). While variations occurred in the C and X bands, the spectral index between these two bands were all inverted, with $\alpha = 0.75$ in 2016 June, $\alpha = 0.53$ in 2018 March, and $\alpha = 0.74$ in 2018 May. We noticed that the new and the archival VLBA observations were done at different frequencies. However, the central frequencies for both observations are very close, actually within the bandwidths. Therefore, the calculated spectral index represents the overall spectral shape well for the studied frequency ranges, i.e., the C–X bands.

As shown in Figure 5, clear variations between our VLBA measurements are present (see also Table 3). Interestingly, our VLBA flux densities, although they exhibit significant variations, are all lower than those of the low-resolution observations. Even the highest VLBA flux density at 8 GHz is still smaller than the VLA measurements. The differences between VLBA and low-resolution flux densities likely arise from either the lost (resolved out) extended emission in the VLBA images, or variability, or both. The possible origin of diffuse/extended emission could be the radio emission from the star formation or the wind, or the jet components/lobes. The largest angular scale our VLBA observations detected is about $12 \times 22$ mas in the S band on 2018 March 30, while that of VLA detection is around $700 \times 100$ mas in the X band on 1998 May 18. As these two observations all show a single compact component, most of the diffuse/extended emission likely lies in between these two angular sizes, which however could be resolved out in high-resolution VLBA observations. We notice that the separation of the companion Seyfert 2 galaxy from our target is about $3''$ (Paliya et al. 2020); thus, any of its radio emission will be included in the low-resolution data. However, any putative contribution of the companion source can be ignored due to the fact that it was not detected at an upper limit of 0.13 mJy from a VLA X band observation made on 1998 May 18 (with a beam size of about $0\rlap.{''}32 \times 0\rlap.{''}23$, Paliya et al. 2020). It is also possible that the reason for the concave spectrum due to the inclusion of nearby sources in low-resolution data at low frequencies. We checked the Faint Images of the Radio Sky at Twenty-Centimeters (FIRST) images, and found that there is only one single FIRST component within $2'$ of the optical position, indicating that there are no strong radio emitters around our target. Moreover, several high-frequency data sets, such as The Parkes-MIT-NRAO (PMN) surveys (Griffith et al. 1995) and The Australia Telescope 20 GHz (AT20G) Survey

---

[10] http://vizier.u-strasbg.fr/viz-bin/VizieR

**Table 4**
Radio Data Collected from VizieR

| $\nu$ (GHz) (1) | $S$ mJy (2) | $\sigma_S$ (mJy) (3) | Observation Date (4) | References (5) | Spatial Resolution (6) |
|---|---|---|---|---|---|
| 0.074 | 750   | 130  | 2003-9-20       | (1)  | 80″ |
| 0.076 | 867.9 | 69.4 | 2013-8-13       | (2)  | ∼100″ |
| 0.084 | 782.1 | 62.6 | 2013-8-13       | (2)  | ∼100″ |
| 0.092 | 652.7 | 52.2 | 2013-8-13       | (2)  | ∼100″ |
| 0.099 | 609.7 | 48.8 | 2013-8-13       | (2)  | ∼100″ |
| 0.107 | 561.9 | 45.0 | 2013-8-13       | (2)  | ∼100″ |
| 0.115 | 463.5 | 37.1 | 2013-8-13       | (2)  | ∼100″ |
| 0.122 | 431.0 | 34.5 | 2013-8-13       | (2)  | ∼100″ |
| 0.130 | 452.4 | 36.2 | 2013-8-13       | (2)  | ∼100″ |
| 0.143 | 404.5 | 32.4 | 2013-8-13       | (2)  | ∼100″ |
| 0.151 | 447.4 | 35.8 | 2013-8-13       | (2)  | ∼100″ |
| 0.158 | 372.4 | 29.8 | 2013-8-13       | (2)  | ∼100″ |
| 0.166 | 389.1 | 31.1 | 2013-8-13       | (2)  | ∼100″ |
| 0.174 | 362.9 | 29.0 | 2013-8-13       | (2)  | ∼100″ |
| 0.181 | 317.9 | 25.4 | 2013-8-13       | (2)  | ∼100″ |
| 0.189 | 319.7 | 25.6 | 2013-8-13       | (2)  | ∼100″ |
| 0.197 | 335.2 | 26.8 | 2013-8-13       | (2)  | ∼100″ |
| 0.204 | 292.8 | 23.4 | 2013-8-13       | (2)  | ∼100″ |
| 0.212 | 288.4 | 23.1 | 2013-8-13       | (2)  | ∼100″ |
| 0.220 | 277.3 | 22.2 | 2013-8-13       | (2)  | ∼100″ |
| 0.227 | 311.2 | 24.9 | 2013-8-13       | (2)  | ∼100″ |
| 0.150 | 383.5 | 39   | 2016-03-15      | (3)  | 25″ |
| 0.365 | 277   | 30   | …               | (4)  | 6″ |
| 1.4   | 98.7  | 20   | 1997-5-26       | (5)  | 5″ |
| 1.4   | 96.1  | 2.9  | 1993-9-20       | (6)  | 45″ |
| 4.85  | 102   | 12   | 1990-11-6       | (7)  | 168″ |
| 8.4   | 97.5  | …    | …               | (8)  | … |
| 8.46  | 91.3  | …    | 1998-5-18       | (9)  | 0″.2 ∼ 0″.04[a] |
| 19.9  | 113   | 6    | 2007-10-(26–30) | (10) | 34″ |

**Notes.** Column (1): observing frequency; Column (2): flux density at $\nu$; Column (3): uncertainty of the flux density; Column (4): observation date; Column (5): references (1) The VLA Low-Frequency Sky Survey (VLSS; Cohen et al. 2007); (2) GLEAM (Hurley-Walker et al. 2017); (3) The Giant Metrewave Radio Telescope (GMRT) 150 MHz All-sky Radio Survey (Intema et al. 2017); (4) The Texas Survey (TXS; Douglas et al. 1996); (5) FIRST (White et al. 1997); (6) The NRAO VLA Sky Survey (NVSS; Condon et al. 1998); (7) PMN (Griffith et al. 1995); (8) CRATES (Healey et al. 2007); (9) NRAO VLA (Archival data); and (10) AT20G survey (Murphy et al. 2010); Column (6): the spatial resolution of the observations.
[a] https://public.nrao.edu/telescopes/vla/

**Table 5**
Power-law Fitting for the Radio Spectrum of J2118-0732

| Resolution (1) | Observation Date (2) | $\alpha_r$ (3) | Note (4) |
|---|---|---|---|
| Low resolution  | 1990–2016     | −0.83 | Low frequency |
| …               | …             | 0.03  | High frequency |
| High resolution | 2016 and 2018 | 0.17  | |
| …               | 2018          | 0.10  | |
| …               | 2018-3-30     | 0.09  | |

**Note.** Column (1): the spatial resolution of the data. VLBA data is high resolution, and others are low resolution (see Table 4); Column (2): observation date; Column (3): the radio spectral index from the power-law fit; Column (4): note for the frequency (see the text).

(Murphy et al. 2010), provide similar to or even worse resolution than the low-frequency ones (e.g., GaLactic and Extragalactic All-sky MWA (GLEAM) survey, Hurley-Walker et al. 2017). The low-frequency observations, for example, GLEAM, are more sensitive to diffuse or extended emission, while the high-frequency observations are better for the detection of the radio core. Therefore, the steep spectrum below 1.4 GHz probably arises from the diffuse/extended emission, while the inverted radio spectrum above 1.4 GHz likely is dominated by the radio core (e.g., Konigl 1981). Alternatively, the steep component may be caused by the interaction between the source and its companion, which triggers a terminal shock (Doi et al. 2018, 2019; Paliya et al. 2020), since no lobe was detected from the FIRST and 1998 VLA images. It is also not impossible that the steep spectrum at low frequencies is due to the relic emission of previous activity, which is especially significant at low frequencies, as found in nearby massive clusters (e.g., A3376, Chibueze et al. 2022). Moreover, we noticed that the emission at the lowest frequencies is from MWA observations. The resolution of MWA is about 100″, corresponding to about 400 kpc for our source; therefore, the radio emission from a several hundred kiloparsec halo, if it exists, will be hardly resolved with MWA observations, and it will then be included in the GLEAM low-frequency emission. The several hundred kiloparsec halo could be a viable alternative for the low-frequency emission. The present low-resolution data preclude us from drawing any firm conclusions. Further observations at low frequencies with higher resolution will be crucial to understanding the radio structure in detail.

The concave spectrum detected in our source was previously found in blazars, for instance, TXS 1700+685 shows steep plus flat radio spectra (Keenan et al. 2021). These two components are thought to be from the extended emission and core, respectively. Moreover, the low-frequency power law and higher-frequency peaked spectra have been observed in a large number of sources at the center of clusters (Hogan et al. 2015). The explanation for the overall spectrum is that the power-law component (low frequency) corresponds to the existence of an extended lobe and a large-scale (several hundred kiloparsec) halo (Baum et al. 1990), while the flux and time variations of the peak component (high frequency) are related to the milliarcsecond-scale components in the jet (Suzuki et al. 2012). Namely, the peak component in the spectrum and the component below the turnover frequency probably represent the latest activities in the core, and older diffuse emission, respectively (Baum et al. 1990; Edwards & Tingay 2004; Torniainen et al. 2007; Hancock et al. 2010). Similarly, all these scenarios may be applicable to the concave spectrum found in J2118-0732.

### 4.3. Comparison with Other γ-Ray NLS1s and Blazars

At present, the number of confirmed γ-ray NLS1s has reached ∼20 (Foschini et al. 2011, 2021; Romano et al. 2018). There are a total of nine NLS1s listed in the Fourth Fermi-LAT source catalog (4FGL; Abdollahi et al. 2020). We compare the radio properties of J2118-0732 with those of the other eight NLS1s. All these sources, SBS 0846+513 (D'Ammando et al. 2012), PKS 2004−447 (Orienti et al. 2012, 2015), 1H 0323 +342 (Wajima et al. 2014; Hada et al. 2018), FBQS J1644+2619 (Doi et al. 2007, 2011), B3 1441+476 and IERS B1303+515 (Gu et al. 2015), PKS 1502+036

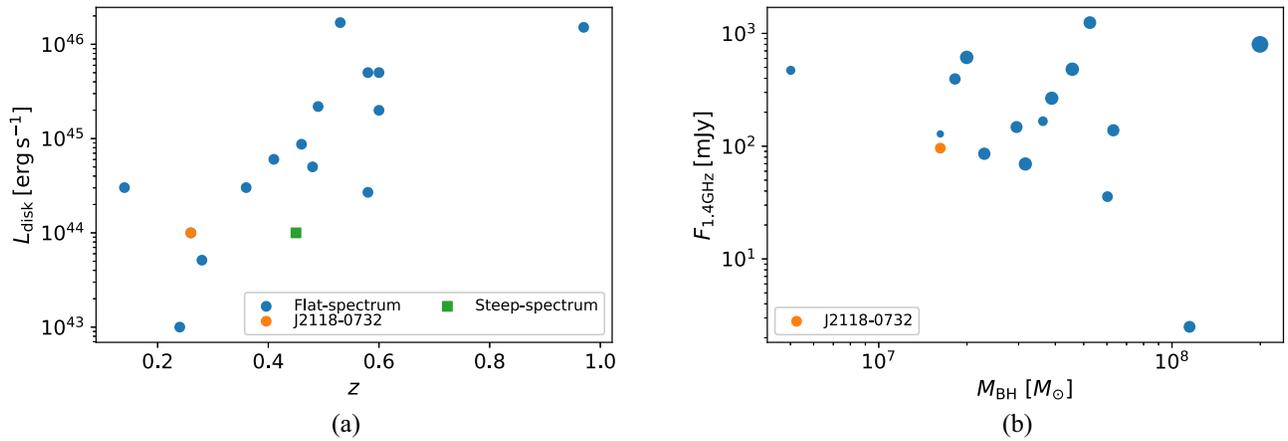

**Figure 6.** The position of J2118-0732 in parameter space ($L_{disk}$ vs. $z$ and $F_{1.4GHz}$ vs. $M_{BH}$). (a) the distribution of the disk luminosity and redshift of the $\gamma$-ray NLS1 sample of Paliya et al. (2019), in which J2118-0732 was classified as a flat-spectrum radio source; (b) the distribution of the flux density at 1.4 GHz and black hole mass for the same sample in panel (a). The size of the circles corresponds to redshift, with a larger size for higher redshift.

(D'Ammando et al. 2013), and PMN J0948+0022 (Giroletti et al. 2011; D'Ammando et al. 2014; Doi et al. 2019) exhibit core-jet structures on a parsec-scale from VLBI observations. Our VLBA observations show that J2118-0732 resembles these objects. The tentative detection of superluminal jet apparent motion (see Section 4.1) in J2118-0732 is also consistent with three $\gamma$-ray NLS1s monitored in MOJAVE (Lister et al. 2016), implying significant beaming effects from the relativistic jet. On the other hand, a subluminal motion was also found in the other two $\gamma$-ray NLS1s from MOJAVE. At the kiloparsec scale, it is interesting that three out of nine $\gamma$-ray NLS1s exhibit two-sided morphologies: 1H 0323+342, PMN J0948+0022, and FBQS J1644+2619 (Antón et al. 2008; Doi et al. 2012, 2019). In addition, IERS B1303+515 shows an extended structure in a Karl G. Jansky Very Large Array (JVLA) image (Berton et al. 2018). Besides these nine sources, most of the additional $\gamma$-ray NLS1s have kiloparsec-scale radio emission (NVSS J095820 +322401, Helfand et al. 2015; NVSS J093241+530633, Singh & Chand 2018; SDSS J110223.38+223920.7, Berton et al. 2018; SDSS J124634.65+023809.0, Gu et al. 2015; Berton et al. 2018; SDSSJ164100.10+345452.7, Berton et al. 2020; 4C +04.42, Lister et al. 2016; and 3C 286, Spencer et al. 1989). NVSS J095820+322401 also has a parsec-scale jet (Linford et al. 2012). 4C +04.42 even has a two-sided jet on the kiloparsec scale, and a one-sided jet on the parsec scale detected by MOJAVE. 3C 286 exhibits the core plus a second lobe (~19.4 kpc) morphology in the VLA image (Spencer et al. 1989), and a core-jet structure at the parsec scale has been detected by VLBA (An et al. 2017). While most sources show the core-jet structure at the parsec scale, a core-only structure was also found in a few objects. 9C J1520+4211 is the point-like source monitored by VLA, VLBA, and Multi-Element Radio-Linked Interferometer Network (MERLIN; Bolton et al. 2006), and NVSS J142106+385522 shows the core structure in VLBA images (Helmboldt et al. 2007). In our source J2118-0732, no kiloparsec-scale emission was detected based on FIRST images and VLA observations at subarcsecond resolutions. From the VLA observation, the size of the compact structure is about $700 \times 1000$ mas, corresponding to $2.82 \times 4.02$ kpc. Tentatively assuming a jet velocity of $0.1c$, and taking a viewing angle of 9°, we could obtain the upper limit of the age of J2118-0732, about 77.5 kyr. However, the strongest extended emission is expected at low frequencies, so further explorations are needed at (sub-)arcsecond resolutions to investigate its presence.

To put the properties of J2118-0732 into context, in Figure 6 we show the position of J2118-0732 in two parameter spaces: disk luminosity against redshift and flux density at 1.4 GHz against black hole mass. These parameters are from the $\gamma$-ray NLS1 sample of Paliya et al. (2019), in which J2118-0732 has been studied; the disk luminosity is calculated by reproducing the emission with a standard disk model and the black hole mass is derived from the single-epoch spectroscopy. We find that J2118-0732 has a rather low disk luminosity ($10^{44}$ erg s$^{-1}$, Paliya et al. 2019), along with a rather low redshift among this sample, see panel (a) in Figure 6. It also has a relatively low black hole mass of $10^{7.21}$ $M_\odot$ and core flux at 1.4 GHz (panel (b)). Chen & Gu (2019) studied a sample consisting of 96 FSRQs, 81 BL Lacs, and 42 flat-spectrum radio-loud narrow-line Seyfert 1 galaxies (F-RLNLS1s). They found that a large number of F-RLNLS1s have a strong radiation-dominated accretion disk and have larger disk luminosity than the jet radiation power. The bolometric luminosity of J2118-0732 is about $6.2 \times 10^{44}$ erg s$^{-1}$ (Yang et al. 2018) calculated assuming $L_{bol} = 9\lambda L_{5100}$, in which the continuum luminosity at 5100 Å is estimated from the broad H$\beta$ line luminosity; thus, the contamination of jet emission is eliminated (see the details in Yang et al. 2018). In contrast, the jet radiative power is about $3.2 \times 10^{43}$ erg s$^{-1}$ (see Section 3.3), and thus is substantially smaller than the bolometric luminosity. As illustrated in Abramowicz et al. (1988), the accretion disk will be a slim disk instead of a standard one when the Eddington ratio is larger than ~0.5. Taking the bolometric luminosity of $6.2 \times 10^{44}$ erg s$^{-1}$, and black hole mass of $10^{7.21}$ $M_\odot$, the Eddington ratio of our source is about 0.3; thus, the spectral energy distribution (SED) revealed by J2118-0732 should not deviate much from that of the standard disk, and the standard disk will still be applicable for SED modeling. A comparison of the broadband parameters derived from the SED modeling performed by Paliya et al. (2019) reveals the similarity of $\gamma$-ray NLS1s with blazars, in particular, with FSRQs. The $\gamma$-ray NLS1s host relatively low-power jets with small bulk Lorentz factors with respect to blazars. The flat-spectrum NLSy1, especially $\gamma$-ray NLS1s, have been proposed to be the low-$M_{BH}$, low-power, counterpart of FSRQs (Foschini et al. 2012).

The VLBA observations have revealed that J2118-0732 shows a high brightness temperature, compact core-jet morphologies, significant variations, and flat spectrum at high frequency. All these properties are similar to those of typical blazars. However, the mild Doppler factor (and jet bulk velocity), and weaker jet power compared to FSRQs indicate that the source is likely a low-$M_{\rm BH}$, low-power version of an FSRQ, and this low power is probably associated with its lower black hole mass (Rakshit et al. 2017; Yang et al. 2018). Several jetted NLS1s have been found to be hosted in merging systems (see, e.g., Olguín-Iglesias et al. 2017; Järvelä et al. 2018; Olguín-Iglesias et al. 2020). A particular feature of J2118-0732 is that it is in an ongoing minor merger system with a companion Seyfert 2 galaxy. Both galaxies have pseudo-bulges and were originally late-type galaxies (Järvelä et al. 2020; Paliya et al. 2020). This is different from the elliptical galaxies commonly found in blazars. Conceivably, our source represents the early evolutionary stage of a blazar, and with the merger ongoing, it could finally evolve into a typical blazar. But the situation could be more complicated in terms of the full set of $\gamma$-ray NLS1s since various types of host galaxies were found. While some sources exhibit either an ongoing merger or traces of violent activities in the past, a few other $\gamma$-ray NLSy1s do not show evidence of mergers and rather support secular-driven black hole growth and jet launching; moreover, a few objects could reside in elliptical galaxies, similarly to most blazars (Paliya 2019).

## 5. Conclusions

We have investigated the radio structure for the $\gamma$-NLS1 J2118-0732 using VLBA and VLA images. The core-jet structure was clearly detected in multifrequency VLBA observations. The VLBA core exhibited slight variations between 2016 and 2018 but significant changes in 2018. High brightness temperatures were found from both VLBA measurements and flux density variabilities. These indicate the presence of a significant beaming effect, though modest compared to typical blazars. The broad radio spectrum of J2118-0732 shows a steep spectrum at low frequencies, but an inverted spectrum at high frequencies. This likely indicates the presence of significant diffuse/extended emission at low frequencies. Considering its many similarities to blazars and its low black hole mass, J2118-0732 is probably the low black hole mass, low-power counterpart of an FSRQ, and may finally evolve into an FSRQ since it is undergoing a merger.

The authors are grateful for Paul Wiita's help with editing the English, and for providing valuable suggestions. We also thank S. Komossa for the suggestion on the classification of NLS1, and M. H. Zhou and J. W. Li for discussions on the data reduction. This work is supported by the National Science Foundation of China (grants 11873073, and 11933007), and the Shanghai Pilot Program for Basic Research at the Chinese Academy of Science, Shanghai Branch (JCYJ-SHFY-2021-013). Support for this work was also provided by the science research grants from the China Manned Space Project under grant No. CMSCSST-2021-A06, and the Original Innovation Program of the Chinese Academy of Sciences (E085021002).

*Facilities:* VLA, VLBA.

*Software:* AIPS (Greisen 2003), DIFMAP (Shepherd 1997).

## ORCID iDs

Xi Shao ● https://orcid.org/0000-0002-4685-4138
Minfeng Gu ● https://orcid.org/0000-0002-4455-6946
Yongjun Chen ● https://orcid.org/0000-0001-5650-6770
Hui Yang ● https://orcid.org/0000-0002-8832-6077
Su Yao ● https://orcid.org/0000-0002-9728-1552
Zhiqiang Shen ● https://orcid.org/0000-0003-3540-8746